\documentclass{elsart5p}

\usepackage{amsmath}
\usepackage{amssymb}
\usepackage{graphicx}
\usepackage{bm}

\newcommand{\mrm}{\mathrm}
\newcommand{\sub}[1]{\mathrm{\scriptscriptstyle{#1}}}

\begin{document}

\begin{frontmatter}

\title{Ultracold-neutron infrastructure for the gravitational spectrometer GRANIT}

\author[ILL,E18]{P. Schmidt-Wellenburg\thanksref{PSW}}
\ead{schmidt-w@ill.fr},
\author[ILL]{K.H. Andersen},
\author[ILL]{P. Courtois},
\author[ILL]{M. Kreuz},
\author[ILL]{S. Mironov}
\author[ILL]{V.V. Nesvizhevsky},
\author[LSPC]{G. Pignol},
\author[LSPC]{K.V. Protasov},
\author[ILL]{T. Soldner},
\author[LSPC]{F. Vezzu},
\author[ILL,E18]{O. Zimmer}

\address[ILL]{Institut Laue Langevin, 6, rue Jules Horowitz, BP-156, 38042 Grenoble Cedex 9, France}
\address[E18]{ Physik-Department E18, Technische Universit\"{a}t M\"{u}nchen, 85748 Garching,
Germany}
\address[LSPC]{Laboratoire de Physique Subatomique et de Cosmology, 53, rue des Martyrs, 38026 Grenoble, France}

\thanks[PSW]{Tel.: +33
4 76 20 70 27; fax: +33 4 76 20 77 77}

\begin{abstract}
The gravitational spectrometer GRANIT will be set up
at the Institut Laue Langevin.
It will profit from the high ultracold neutron density produced by
a dedicated source.
A monochromator made of crystals 
from graphite intercalated with potassium will provide a neutron beam with
$8.9$~\AA\, incident on the source.
The source employs superthermal conversion of cold neutrons in superfluid
helium, in a vessel made from BeO ceramics with Be windows.
A special extraction technique has been tested which feeds the spectrometer only with neutrons 
with a vertical velocity component $v_{\bot}\leq 20$~cm/s, thus keeping the density in the source high. 
This new source is expected to provide a density of up to $\rho = 800~\mrm{cm^{-3}}$ for the spectrometer.
\end{abstract}


\end{frontmatter}

\section*{Introduction}
The solutions for Schr\"{o}dinger's equation for a neutron bouncing on a
reflecting horizontal surface in the Earth's gravitational field are given by
Airy functions\,\cite{flugge1971,Luschikov1978}. This textbook example of bound energy
states in a linear potential has been demonstrated experimentally at the high
flux reactor of the Institut Laue
Langevin\,\cite{Nesvizhevsky2002,Nesvizhevsky2003}. A new gravitational
spectrometer, GRANIT\,\cite{Nesvizhevsky,Granit2008}, is being built to
investigate these quantum states further, and to induce resonant
transitions between gravitationally bound quantum states. Proposed applications are
a refined measurement of the electrical charge of the neutron, the search
for the axion\,\cite{Baessler2007,Axion2008} and other additional forces
beyond the standard model.
Experiments at ILL's present facility PF2 for ultracold neutrons (UCN) are limited by counting
statistics and systematic effects\,\cite{Pignol2007}. In this paper we describe design and 
status of a dedicated UCN source and its coupling to the gravitational
spectrometer, updating a previous status report \cite{Schmidt-Wellenburg2007a}.

\section*{General aspects of Helium-4 based superthermal converters}
The UCN source employs down-conversion of monochromatic cold neutrons with wavelength around
8.9~\AA\ in superfluid helium, via single-phonon excitation \cite{Golub1975}.
The saturated UCN density in the converter,

\begin{equation}
	\rho_{\mathrm{\scriptscriptstyle{UCN}}}=P \cdot \tau,	
\end{equation}
\label{eq:density}

\noindent is determined by production rate density $P$ and neutron storage time constant $\tau$. The rate

\begin{equation}
	1/\tau = \frac{1}{\tau_{\beta}}+\frac{1}{\tau_{\sub{wall}}}+\frac{1}{\tau_{\sub{up}}}+\frac{1}{\tau_{\sub{abs}}}+\frac{1}{\tau_{\sub{extr}}},
\end{equation}
\label{eq:lifetime}

\noindent includes contributions from all existing loss channels (neutron
beta decay, losses due to wall collisions, up-scattering due to phonons,
nuclear absorption by $ ^3$He\,-\,impurities in the helium, extraction of
UCN\footnote{Helium-4 based UCN sources are usually used in accumulation mode,
without extraction during accumulation. In this paper, we are discussing a
source providing a quasi-continuous flux of UCN.}).

As the absorption cross section of $ ^4$He is zero there is no absorption
inside a pure converter.
For $T \longrightarrow 0$~K
the up-scattering cross section becomes negligible, and therefore $\tau$
is determined only by neutron beta decay, losses due to wall collisions,
and extraction of UCN.

The UCN production rate density
is defined as the conversion rate of neutrons to energies below the Fermi
potential of the walls of the converter volume. For beryllium with
$V_{\mrm{F}}=252$~neV
the production rate density due to the single-phonon process, calculated from
neutron scattering data, is 
$P_{\mrm{I}}=(4.97\pm 0.38) \cdot 10^{-8} \mrm{d}\Phi/\mrm{d}\lambda|_{\lambda^{\ast}} \mrm{s^{-1}cm^{-3}}$,
with the differential flux at $\lambda^{\ast} = 8.9$~\AA~given in
$\mrm{cm^{-2}s^{-1}}$\,\AA\,$^{-1}$\,\cite{Schmidt-Wellenburg2008}.

\section*{Source concept}
\begin{figure}
	\centering
	\includegraphics[width=0.8\columnwidth]{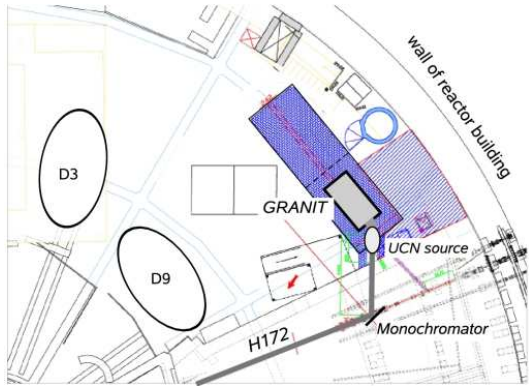}
	\caption{Setting of the monochromator, the source, and GRANIT inside level C of the ILL.}
        \label{fig:position}
\end{figure}

The UCN source employs cold neutrons from the neutron beam H172 on
level C of the high flux reactor at the ILL\,(see fig.\,\ref{fig:position}).
A crystal monochromator, positioned 12~m downstream from the liquid-deuterium 
cold source situated in-pile reflects neutrons with 8.9~\AA\ out of
the direct beam. It feeds a secondary, 4.5~m long neutron guide
equipped with $m=2$ supermirror coatings.
The superfluid-helium UCN converter with its special UCN extraction system is installed at the end of this guide. 
The individual components of the source implementation are described in the following subsections. 

\subsection*{Neutron monochromator}
Single-phonon downscattering of cold neutrons in superfluid helium as UCN
production mechanism requires only a narrow range of wavelengths around
8.9~\AA\,. Using a monochromator and a secondary guide strongly reduces
the background with respect to a converter placed in the direct, white beam.
The price to be paid is a reduction in intensity due to imperfection of
monochromator and secondary guide, and due to omission of multi-phonon
processes\,\cite{Baker2003,Schmidt-Wellenburg2008}.

Crystal monochromators reflect neutrons of the desired wavelengths away from the primary beam under the $d$-spacing dependent Bragg angle:

\begin{equation}
	\theta_{\mrm{B}}=\arcsin{\left(\frac{n\lambda}{2d}\right)}.
\end{equation}

\noindent The $d$-spacing of a monochromator for $\lambda=8.9$~\AA\, has to be $d>\lambda/2 = 4.45$~\AA. In a perfect single crystal the line width for Bragg reflected neutrons is extremely small, $\Delta k/k = 10^{-4}$\,\cite{Goldberger1947} leading to a very narrow acceptance angle of incident neutrons:

\begin{equation}
	\frac{\Delta k}{k} = \cot{\theta \Delta \theta}.
\end{equation}

\begin{figure}
	\centering
	\includegraphics[width=0.8\columnwidth]{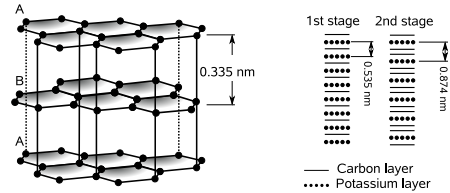}
	\caption{Crystal structure of graphite and staging of GiC. The intercalant diffuses into the graphite and thus increases the lattice spacing. Different stages of GiC can be produced. The stage number refers to the number of unperturbed graphite layers between two layers of intercalant atoms.}
	\label{fig:gic}
\end{figure}

\noindent The incident beam has a divergence of typically $\pm 2^{\circ}$ at
$8.9$~\AA\, due to the $m=2$ supermirrors used in the guide. Therefore a
``mosaic crystal''\,\cite{Goldberger1947} is used. Such a crystal can be
regarded as a collection of microscopically small perfect crystals with
differing angles $\epsilon$ with respect to the overall crystal orientation.
Although the angular distribution is in general arbitrary, it resembles a
cylindrically symmetric Gaussian distribution with a width $\eta$, called
the mosaicity. To obtain a high acceptance the mosaicity should be in the
range of the divergence of the incident beam\,\cite{Liss1994}. A high
reflectivity of the crystal for $8.9$~\AA\, neutrons is required, too. 
Mica and graphite intercalated compounds\,(GiC) with alkali metals both match the
requirements for the lattice spacing. However, due to the too small mosaicity
$\approx 0.3^{\circ}$ of mica we are using a potassium intercalated
graphite monochromator of the type already employed at the ILL and
NIST\,\cite{Boeuf1983,DB21,Mattoni2004}.  
For highly oriented pyrolytic graphite the spacing is $d=3.35$~\AA\, with a
typical mosaicity of $1^{\circ} - 2^{\circ}$. The $d$-spacing is increased by placing
guest species in between graphite layers (see fig.\,\ref{fig:gic}). This
process is called intercalation.

Alkali intercalated graphite compounds are conveniently produced using the
``two-bulb'' technique, where the graphite is maintained at a temperature
$T_{\mrm{c}}$ which is higher than $T_{\mrm{a}}$ of the alkali
metal\,\cite{Herold1955}. The stage which is formed depends on the
temperature difference $\Delta T =  T_{\mrm{c}}-T_{\mrm{a}}$ and the
quantity of potassium available. For stage-1, stage-2 we are employing
5\,g, 1\,g~ampoules of potassium at $T_{\mrm{a}}=255\pm 3~^{\circ}\mrm{C}$
and $\Delta T_1 = 10\pm 3~^{\circ}\mrm{C}$, $\Delta T_2 = 102\pm 3~^{\circ}\mrm{C}$,
respectively. 

\begin{figure}
	\centering
	\includegraphics[width=0.8\columnwidth]{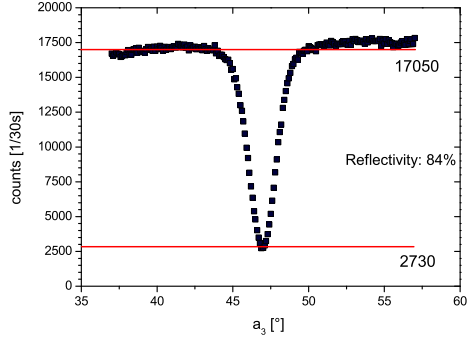}
	\caption{Reflectivity measured in transmission with $5.9$~\AA\, neutrons on the three axis spectrometer IN12 of the Insitut Laue Langevin.}
	\label{fig:refl}
\end{figure}

The monochromator consists of 18 stage-2
potassium intercalated graphite crystals ($\mrm{C_{24}K}$)
providing a lattice spacing of
$d=8.74$~\AA\, giving a take-off angle of $2\theta = 61.2^{\circ}$. The
typical mosaic spread of the produced crystals is
$\eta = 1.5^{\circ} - 2.2^{\circ}$ which matches the incident
divergence of beam H172.
Furthermore
the thermal-neutron absorption cross sections of carbon (0.0035~barn) and potassium
(2.1~barn) are small. The crystals are mounted pair-wise onto
graphite bars which then are screwed into an indium sealed aluminum box. 
First crystals with a reflectivity of
$r\geq 80$~\% (see fig.\,\ref{fig:refl}) have been produced.

A second monochromator with a take-off angle of $2\theta = 112.5^{\circ}$
made of a set of stage-1 crystals (C$_{8}$K, $d=5.35$~\AA) will be placed
close to the first one on a rotary table. The two monochromators
can be interchanged making two separate $8.9$~\AA~ beam ports available at H172
(see fig.\,\ref{fig:position}). The second beam will feed a position
for further tests and developments on liquid helium based
UCN sources, and later for the cryo-EDM experiment\,\cite{vdGrinten2008}. 

\subsection*{Secondary neutron guide, UCN converter and cryostat}
General requirements for producing a high density of UCN in a converter vessel filled with superfluid-helium
are an intense incident beam, and for the walls a high Fermi potential with a small loss-per-bounce coefficient $\mu$. A converter 
vessel with polished, flat wall surfaces may keep together a divergent incident beam. On the other hand, for short sources,
rough surfaces seem to support faster UCN extraction compared to polished ones\,\cite{Zimmer2008}. 
A converging secondary neutron guide may increase the flux density of 8.9~\AA\, neutrons at the entrance to the 
converter. However, the increased divergence might result in a strong decrease of flux density along the converter for 
rough and/or low-Fermi-potential walls.

\begin{figure}
	\centering
	\includegraphics[width=0.9\columnwidth]{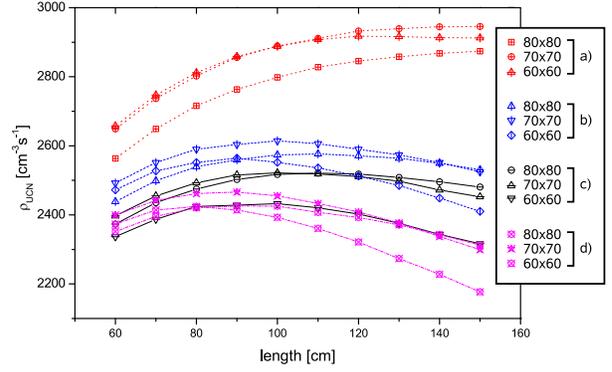}
	\caption{Simulation of UCN density as a function of length and cross
	  section of the conversion volume, for four different combinations
	  of wall coatings of guide and conversion volume: a)\,m=2 side
	  walls and m=3 top and bottom wall of guide, m=2 conversion volume;
	  b)\,m=3 guide, m=1 conversion volume; c)\,m=3 guide,
	  m=0 conversion volume; d)\,m=2 side walls and m=3 top and
	  bottom wall of guide, m=0 conversion volume. The beam after the
	  monochromator is taken to have a Gaussian divergence
	  $\alpha = 1.8^\circ$ and a differential flux of
	  $\mrm{d}\phi/\mrm{d}\lambda |_{\lambda^{\ast}}=3\cdot10^8~\mrm{cm^{-2}s^{-1}}$\,\AA\,$^{-1}$.
	  The higher the m-value the higher the density, but the gain for a
	  volume of 100~cm length is negligible. However, the combination of
	  a high m-value and an increased length of the conversion volume
	  increases the density.} 
		\label{fig:density_sim}
\end{figure}

Monte Carlo simulations were performed to investigate UCN production for various guide geometries and 
wall properties (see fig.\,\ref{fig:density_sim}). Taking into account also other constraints we decided to employ as converter vessel an assembly of 5 rectangular BeO tubes, closed by Be foils, thus taking advantage of the high Fermi potential 
of these materials (261~neV and 252~neV, respectively). The flux incident on the monochromator,
$\mrm{d}\phi/\mrm{d}\lambda|_{\lambda^{\ast}}=6\cdot 10^8 \mrm{s^{-1}cm^{-3}}$,
was calculated from known cold-source data and a transmission simulation
for the existing guide. The peak reflectivity of  $\geq80$~\% of the
monochromator corresponds to an integral reflectivity of $\approx 50$~\%
for the divergent 8.9~\AA\, beam. The neutron guide between
monochromator and source converges over a length of
$4.5$~m from a $80\times 80 \mrm{mm^2}$ to
the cross section indicated in fig.\,\ref{fig:density_sim}. 

The simulations show that, for a conversion volume with the properties
defined above, with a cross section of 70$\times$70~mm$^2$, a length of 1~m,
and taking $\mu=1\cdot10^{-4}$, 
we expect an UCN density of $\rho_{\mrm{calc}}\geq 2500~\mrm{cm^{-3}}$.

\begin{figure}
	\centering
	\includegraphics[width=\columnwidth]{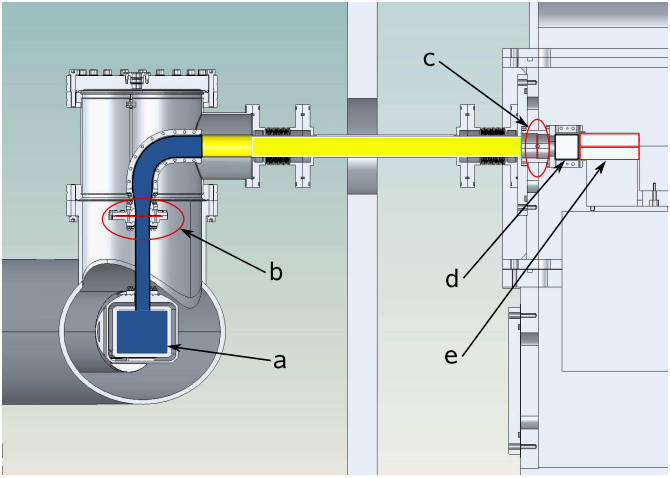}
	\caption{Drawing of the UCN extraction system: a -- UCN production
	  volume (cut), b -- cold UCN shutter, c -- room temperature UCN
	  shutter, d -- intermediate volume, e -- semidiffuse channel.
	  The cold part of the extraction system is marked in blue, the
	  room-temperature part in yellow.}
	\label{fig:extraction}
\end{figure}

UCN are extracted from the converter through specular tubes
into an intermediate volume (see fig.\,\ref{fig:extraction}).
To suppress heat load from thermal radiation 
along the tube we use an UCN shutter thermally anchored to the $4$~K screen of the cryostat. The
intermediate volume, at room temperature, is equipped with a second
UCN shutter at the entrance. Both shutters are operated in sequential
mode to establish and maintain a high UCN density inside the intermediate volume.
The UCN yield of our window-free, vertical UCN extraction is much improved with respect to an older scheme with horizontal
UCN extraction through foils\,\cite{Kilvington1987}. A bent in the extraction guide reduces background from directly scattered
cold neutrons.

For filling and cooling the converter volume with liquid helium we
employ a cryostat previously developed at Munich. For some general description 
of the apparatus and first experiments on UCN production see\,\cite{Zimmer2007}. 
The apparatus includes a Gifford McMahon cryocooler with 1.5~W at 4.2~K, which serves for
liquefaction of helium from gas bottles\,\cite{Schmidt-Wellenburg2006} and for cooling heat screens to 50~K and 4~K. 
A $ ^4$He evaporation stage cools liquid helium to the superfluid phase,
allowing us to use a superleak to remove $ ^3$He. The cooling of
the converter volume is achieved with a $ ^3$He closed cycle evaporation
stage. Using a roots blower pump with 500~$\mrm{m^3/h}$ nominal pumping
speed backed by a 40~$\mrm{m^3/h}$ multiroots pump we were so far able
to cool the filled converter to 0.7~K. An ongoing upgrade of the apparatus shall bring the temperature down to 0.5~K.

\subsection*{UCN selection with semidiffuse channel}
The first experiments with the gravitational spectrometer GRANIT require a quasi-continuous flux of UCN within a narrow phase space element, 
for which a special UCN extraction system has been developed. 
UCN from the converter are guided to the intermediate volume by
a highly polished nickel or diamond-like carbon\,(DLC)
coated guide (fig.\,\ref{fig:extraction}). The intermediate volume is made
of rough DLC-coated
aluminum plates, providing mixing in the phase space of stored neutrons.
The optimum size of the volume was
determined by simulations with Geant4UCN\,\cite{atchinson2005} to be
40$\times$40~$\mrm{mm^2}$, the width 300~mm is given by the dimensions
of the spectrometer. 

\begin{figure}
	\centering
	\includegraphics[width=0.9\columnwidth]{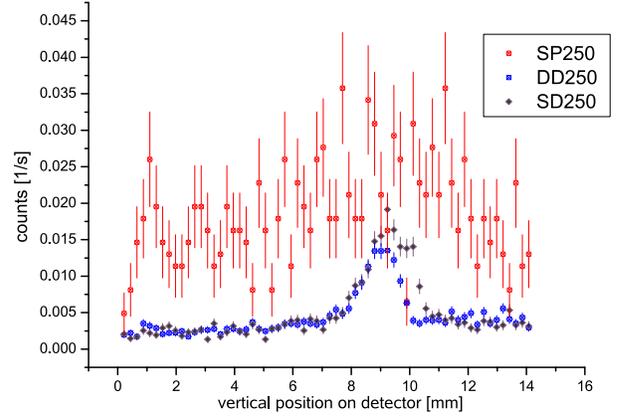}
	\caption{Vertical distribution of transmitted neutrons after an
  extraction channel in different configurations (two mirror like surfaces
  \textbf{SP}, mirror on bottom and rough surface on top \textbf{SD}, and
  two rough surfaces \textbf{DD}). The channel height was $h=250~\mrm{\mu m}$.
  The distributions were measured with a high resolution UCN detector
  (pixel size:\,$55 \times 55 \mrm{\mu m^2}$)\,\cite{Jakubek2008},
  placed $\approx 60$~mm behind the exit of the channel, with position 0
  about 9~mm below the channel. The integral transmission for the SP
  configuration is a
  factor 5 bigger than for the DD configuration. In the SD configuration,
  all desired quantum states can pass into the spectrometer. Configurations
  SD and DD collimate the UCN, visible in the peaks (that are washed out by gravity).}
	\label{fig:trans}
\end{figure}

UCN are extracted from the intermediate volume via a narrow horizontal semidiffuse extraction channel
described in \cite{Schmidt-Wellenburg2007,Barnard2008}. Measurements
have shown that such a channel increases the storage time inside the 
intermediate volume since more than 80\% of the neutrons incident on the 
channel are reflected back. These neutrons are lost in conventional
collimation systems whereas they increase the UCN density here and in turn 
the flux of extracted neutrons.
The phase space selectivity of the extraction channel is demonstrated in fig.\,\ref{fig:trans}.
For GRANIT it will be made of DLC coated quartz plates, with
the lower surfaces polished and the upper ones rough. The channel dimensions
are $h=200~\mrm{\mu m},\,l=100~$mm, $w\,=\,300~$mm, thus selecting neutrons with a vertical energy component $E_{\bot}=\hbar^2p_{\bot}^2/2m \le 20$~peV with respect to the channel bottom. We expect a reflectivity (fraction of neutrons that return from the channel to the storage volume) of $r\geq 80$~\% for all other neutrons.

\section*{Conclusion}
Prototypes of monochromator crystals, the converter, and the semidiffuse
channel have all been tested separately.
The design of the full set-up is finalized, and the components are being
produced. Calculations with an incident differential flux
$\mrm{d}\phi/\mrm{d}\lambda|_{\lambda^{\ast}} = 6\cdot 10^8~\mrm{s^{-1}cm^{-3}}$
on the monochromator give an UCN density of
$\rho_{\mrm{\scriptscriptstyle{UCN}}} \approx 2500~\mrm{cm^{-3}}$ in the
converter and up to $\rho_{\mrm{\scriptscriptstyle{int}}} = 800~\mrm{cm^{-3}}$ in
the intermediate volume. This yields an available phase-space density of
$\Gamma_{\mrm{\scriptscriptstyle{He}}} \approx 0.2~\mrm{cm^{-3}(m/s)^{-3}}$
for the critical velocity of $7$~m/s of the used materials.
Compared to the phase-space density
$\Gamma_{\mrm{\scriptscriptstyle{Turbine}}} \approx 0.013~\mrm{cm^{-3}(m/s)^{-3}}$
of the UCN turbine at the ILL this is more than a factor ten of
improvement.
This calculation assumes perfect conditions and in all parts optimal
transmissions.
The integration of source, intermediate volume and extraction channel
is a challenging task, and further optimization may be needed to approach
the calculated densities.

\section*{Acknowledgement}
We are grateful to our colleagues from the GRANIT collaboration, the monochromator collaboration with NIST, and the DPT of the ILL for fruitful discussions and support. 
This work is supported by the German BMBF (contract number 06MT250) and by the French Agence de la Recherche (ANR).

\end{document}